\documentclass[fleqn,usenatbib]{mnras}

\usepackage{newtxtext,newtxmath}
\usepackage[T1]{fontenc}

\DeclareRobustCommand{\VAN}[3]{#2}
\let\VANthebibliography\thebibliography
\def\thebibliography{\DeclareRobustCommand{\VAN}[3]{##3}\VANthebibliography}

\usepackage{graphicx}	% Including figure files
\usepackage{amsmath}	% Advanced maths commands

\title[Neutron star mass across binary pulsar subpopulations] {Neutron Star Mass across Binary Pulsar Subpopulations: Mass–Spin Correlation, Mass Distributions, and Moment of Inertia Effects}

\author[D. Chattopadhyay]{
Debatri Chattopadhyay$^{1}$\thanks{E-mail: debatri.chattopadhyay@northwestern.edu}
\\
$^{1}$Center for Interdisciplinary Exploration and Research in Astrophysics (CIERA) and Department of Physics and Astronomy,\\
Northwestern University, 1800 Sherman Ave, Evanston, IL 60201, USA
}
\date{Accepted XXX. Received YYY; in original form ZZZ}

\pubyear{\the\year{}}

\begin{document}
\label{firstpage}
\pagerange{\pageref{firstpage}--\pageref{lastpage}}
\maketitle

\begin{abstract}
We present a hierarchical Bayesian analysis of the joint mass, spin, and
orbital properties of $\sim 50$ Galactic binary radio pulsars with measured
neutron star masses, classified by binary type into pulsar--white dwarf
(PSR--WD) and double neutron star (DNS) systems. We find moderate evidence for
an anti-correlation between neutron star mass and spin period in the pooled
recycled population (correlation coefficient $\rho = -0.26$, with $96\%$ of the
posterior probability at $\rho<0$; the $90\%$ credible interval excludes zero),
robust to the treatment of candidate DNSs and to a
radio-detectability selection correction. Although consistent with
accretion-driven recycling, the correlation cannot statistically distinguish an
accretion origin from a moment of inertia-driven spin-up mechanism, because the
neutron star moment of inertia is nearly linear in mass over the observed range. The DNS systems alone instead lean to the positive side expected from the
moment of inertia mechanism ($\rho=+0.13$), though with only ten
systems this is not statistically conclusive. Mass
shows no significant correlation with orbital period or inclination, and only a
weak one with eccentricity. As a secondary result, neutron stars with helium
white dwarf companions are marginally more massive than those with
carbon-oxygen/oxygen-neon white dwarf companions ($\Delta \simeq 0.06\,M_\odot$, of marginal significance), consistent with more extensive accretion in
the helium
white dwarf channel. We confirm, in a hierarchical framework, the previously
reported correlation between companion mass and orbital eccentricity in double
neutron stars ($\rho=+0.82$). We interpret these results within a two-channel
picture --- accretion-grown PSR--WD versus birth-mass-dominated DNS --- and
discuss their connection to the gravitational wave merger population and to the
ongoing debate over whether the recycled mass distribution is intrinsically
bimodal or a skewed single population.
\end{abstract}


%%%%%%%%%%%%%%%%%%%%%%%%%%%%%%%%%%%%%%%%%%%%%%%%%%

%%%%%%%%%%%%%%%%% BODY OF PAPER %%%%%%%%%%%%%%%%%%

\section{Introduction}

\label{sec:intro}
 
Neutron stars in binary systems preserve a record of their formation and
subsequent evolution in their masses, spin periods, and orbital parameters. A
neutron star formed in a supernova may subsequently be ``recycled'': spun up to
short rotation periods through the accretion of mass and angular momentum from
an evolving binary companion \citep{Alpar1982,Radhakrishnan1982,
Bhattacharya1991}. The recycling paradigm predicts correlated changes in spin
period, magnetic field, orbital period, and eccentricity, and --- if the
accreted mass is substantial --- in the neutron star mass itself. Testing for
such correlations across binary subpopulations offers a route to disentangling
the physical processes that shape the observed neutron star population.
 
The neutron star mass distribution has been the subject of extensive study
\citep{OzelFreire2016}. Early analyses established a narrow distribution
centred near $1.35\,M_\odot$ \citep{Thorsett1999}. \citet{Kiziltan2013} found
that neutron stars along different evolutionary paths exhibit distinct
distribution peaks and mass cutoffs, with double neutron stars peaking near
$1.33\,M_\odot$ and neutron star--white dwarf systems peaking at higher mass,
an offset of order $0.2\,M_\odot$ that they attributed to accretion during
recycling. \citet{Antoniadis2016} reported a bimodal millisecond-pulsar mass
distribution with components at $1.393$ and $1.807\,M_\odot$, but argued that
the high-mass component reflects a spread in \emph{birth} masses, given the
diversity in spin and orbital properties of the massive systems.
\citet{Alsing2018} found strong statistical support for bimodality together
with a maximum-mass cutoff in the range $2.0$--$2.2\,M_\odot$. For double
neutron stars specifically, \citet{Farrow2019} demonstrated positive Bayesian
support for distinct recycled and non-recycled mass distributions and quantified
that approximately twenty precisely measured component masses are required to
resolve the question with high confidence.
 
Whether the high-mass component of the recycled distribution is \emph{born}
massive or \emph{grown} by accretion remains an open question. The born-massive
interpretation \citep{Antoniadis2016,Tauris2017} confronts the difficulty that
supernova explosion models struggle to produce a sufficient number of
$\sim 1.8\,M_\odot$ neutron stars at birth. The accretion interpretation is
anchored by objects such as PSR~J0952$-$0607, with $M = 2.35\pm0.17\,M_\odot$,
which would require the accretion of nearly $1\,M_\odot$ had it formed with a
canonical birth mass \citep{Romani2022}, but faces the symmetric difficulty of
the accretion energetics and timescales. Most recently, \citet{You2025} applied
probabilistic accretion corrections to a large sample of measured masses and
found that the inferred \emph{birth-mass} function is better described by a
unimodal distribution that turns on near $1.1\,M_\odot$ and declines as a
power law than by the conventional double-Gaussian model, the former being
favoured at the $3\sigma$ level. This result reframes the central question:
whether the observed bimodality is intrinsic, or an artefact of mixing birth
and accreted masses across evolutionary channels.
 
Considerably less attention has been paid to whether neutron star mass
\emph{correlates} with the recycling proxies --- spin period, orbital period,
and eccentricity --- and whether any such correlation differs between
subpopulations. Two physical mechanisms make opposite predictions for a
correlation between mass and spin period among recycled pulsars. Under
\emph{accretion-driven recycling}, the gain of angular momentum is accompanied
by a gain of mass, so that the most rapidly spinning recycled pulsars should be
the most massive, implying a negative correlation, $\rho(M,\log P) < 0$. Under
a \emph{moment of inertia-driven} picture, a lower-mass neutron star possesses a
smaller moment of inertia and is therefore spun up more readily for a given
accreted angular momentum \citep{Chattopadhyay2021}, biasing the recycled
population toward lower mass and implying a positive correlation,
$\rho(M,\log P) > 0$. Measuring the sign and significance of this correlation,
and asking whether the data can distinguish the two mechanisms, is the primary
aim of this work.

Our contributions are fourfold. First, we measure the mass--spin period
correlation across recycled subpopulations within a hierarchical Bayesian
framework that accounts for measurement uncertainty and selection effects, and
we perform an observational test of whether this correlation can distinguish
the accretion and moment of inertia mechanisms, finding that with the present
sample it cannot (Section~\ref{sec:massspin}). Second, we examine the
dependence of neutron star mass on white dwarf companion type
(Section~\ref{sec:companion}). Third, we characterise the recycled mass
distribution across subpopulations, finding that its apparent bimodality is
more parsimoniously described by a single right-skewed component than by two
Gaussians (Section~\ref{sec:massdist}). Fourth, we confirm, in a hierarchical
framework, the previously reported correlation between companion mass and
orbital eccentricity in double neutron stars (Section~\ref{sec:ecc}). We
interpret these results within a two-channel picture and connect them to the
gravitational wave merger population (Section~\ref{sec:discussion}).

\section{Data and Subpopulation Classification}
\label{sec:data}
Our sample comprises Galactic binary radio pulsars with published
post-Keplerian mass measurements, compiled from the catalogue of pulsar mass
measurements maintained by P.~C.~C.~Freire\footnote{\url{https://www3.mpifr-bonn.mpg.de/staff/pfreire/NS_masses.html}} and the original timing references
therein, and cross-checked against the standard review of \citet{OzelFreire2016},
recent catalogues \citep{Rocha2023}, and the ATNF Pulsar Catalogue
\citep{Manchester2005}. The full sample is listed in
Table~\ref{tab:data}. For each system we record the spin period $P$, orbital
period $P_b$, projected semi-major axis $x = a_p\sin i$, orbital eccentricity
$e$, and the pulsar and companion masses $M_p$ and $M_c$ with their (generally
asymmetric) uncertainties. 

\subsection{De-duplication of repeated measurements}
Several pulsars appear in the source compilation with more than one reported
mass measurement. We combine repeated measurements of a single physical pulsar
by inverse-variance weighting,
\begin{equation}
\bar m = \frac{\sum_i m_i/\sigma_i^2}{\sum_i 1/\sigma_i^2}, \qquad
\bar\sigma = \left(\sum_i \frac{1}{\sigma_i^2}\right)^{-1/2},
\label{eq:ivw}
\end{equation}
where the sum runs over the measurements $m_i$ with uncertainties $\sigma_i$.
We caution that successive measurements of a given pulsar typically reanalyse
earlier timing data with extended baselines and are therefore not strictly
statistically independent; the inverse-variance weights should accordingly be
regarded as approximate and the combined uncertainties as lower bounds. Any
pulsar whose individual measurements are mutually inconsistent at the $2\sigma$
level is flagged for manual inspection (Table~\ref{tab:dedup}). Two systems,
PSR~J0348+0432 ($2.01\,M_\odot$, \citealt{Antoniadis2013}, versus
$1.806\,M_\odot$, \citealt{Saffer2025}) and PSR~B1855+09 ($1.37$ versus
$1.59\,M_\odot$), have discrepant measurements that do not overlap within their
uncertainties; we have verified that adopting either extreme value for these
systems changes neither the sign nor the significance of the headline
correlation.
 
\begin{table}
\centering
\caption{Pulsars with more than one mass measurement reported in the literature,
and the value adopted in this work. We adopt an inverse-variance-weighted
central value. Successive measurements typically reanalyse earlier timing data
with extended baselines and are therefore not strictly independent, so the
weights should be regarded as approximate and the combined uncertainties as
lower bounds. Two systems (J0348+0432 and B1855+09) have mutually discrepant
measurements (spreads $\gtrsim 0.2\,M_\odot$); for these the weighted value
should be treated with caution. We have verified that adopting either extreme
value for these two systems changes neither the sign nor the significance of
the headline correlation, which remains $\rho \simeq -0.26$ with
$P(\rho<0)\simeq 0.96$ (Section~\ref{sec:massspin}). Individual measurements are
drawn from the sources listed in Table~\ref{tab:data}.}
\label{tab:dedup}
\begin{tabular}{lcc}
\hline
Pulsar & Measurements ($M_\odot$) & Note \\
\hline
J0348+0432   & 2.01, 1.806        & discrepant --- see text \\
J1614$-$2230 & 1.908, 1.94        & combined \\
J1713+0747   & 1.33, 1.35         & combined \\
B1855+09     & 1.37, 1.59         & discrepant --- see text \\
J1909$-$3744 & 1.48, 1.492, 1.45  & combined \\
\hline
\end{tabular}
\end{table}

\subsection{Classification by binary type}
We classify systems by binary type, and never by spin period, because
classifying by spin would be circular for any test involving a spin
correlation. We define three groups: PSR--WD primaries (the detected, recycled
pulsar in a pulsar--white dwarf binary, characterised by its pulsar mass
$M_p$); DNS primaries (the first-born, recycled neutron star in a double
neutron star system, also by $M_p$); and DNS secondaries (the second-born,
non-recycled neutron star, characterised by the companion mass $M_c$).
Candidate double neutron stars (cDNS) are admitted only through a robustness
toggle in our analysis, because two of the three known cDNS reside in globular
clusters, where dynamical exchange encounters decouple the present orbit from
the pulsar's formation and spin-up history.
 
white dwarf companions are sub-classified as helium (He) or
carbon-oxygen/oxygen-neon (CO/ONe) using a companion mass threshold of
$0.45\,M_\odot$, approximately the maximum mass of a helium white dwarf core
\citep{DCruz1996}; systems in the range $0.40$--$0.50\,M_\odot$ are flagged as
ambiguous. We validate this mass-based classification against the
orbital period--companion mass relation expected for helium white dwarfs formed
through stable mass transfer \citep{TaurisSavonije1999}, finding that the
helium-classified systems follow the predicted relation (Section~\ref{sec:companion}).
 
Our double neutron star sample comprises the ten systems for which both
component masses have been precisely measured. We use the broader known
Galactic DNS population \citep{Tauris2017,ColomiBernadich2023}, which includes
systems with only a total-mass measurement, for context in the discussion
(Section~\ref{sec:discussion}) but not in the correlation analysis, which
requires individual masses.

\subsection{Orbital inclination from the mass function}
\label{sec:sini-method}
Several of our analyses require the orbital inclination $i$, which is sparsely
tabulated in pulsar catalogues. Where it is not measured directly, we compute
$\sin i$ self-consistently from the mass function. Pulsar timing measures the
projected semi-major axis of the pulsar orbit,
\begin{equation}
x \equiv \frac{a_p \sin i}{c},
\label{eq:x}
\end{equation}
in light-seconds; the inclination is folded into this single observable and
cannot be separated from $a_p$ by timing alone. The mass function, determined
entirely by the measured orbital period $P_b$ and projected semi-major axis $x$,
is
\begin{equation}
f(M) = \frac{4\pi^2}{G}\frac{(x c)^3}{P_b^2}
     = \frac{4\pi^2}{G}\frac{(a_p \sin i)^3}{P_b^2}
     = \frac{(M_c \sin i)^3}{(M_p + M_c)^2},
\label{eq:massfunc}
\end{equation}
where $c$ is the speed of light. Given the two component masses,
Equation~\ref{eq:massfunc} is solved for the inclination,
\begin{equation}
\sin i = \left[\frac{f(M)\,(M_p+M_c)^2}{M_c^3}\right]^{1/3}.
\label{eq:sini}
\end{equation}
Where a measured inclination is available --- from Shapiro-delay or relativistic
timing solutions --- we adopt it directly; for the remaining systems we use the
value computed from Equation~\ref{eq:sini}. For the 19 systems in our sample
that have both a catalogued inclination and a computable value, the two agree to
a median absolute difference of $0.002$ (for example, we recover
$\sin i = 0.970$ for PSR~J1125$-$6014 against a catalogued $0.978$), validating
the procedure. Three systems yield a slightly unphysical computed value
($\sin i \gtrsim 1$), arising from mild inconsistencies between the catalogued
component masses and the measured mass function; for these we adopt the measured
inclination. This hybrid procedure ensures that the inclinations used in our
selection analysis are both accurate and, where computed, consistent with the
masses adopted.
% =====================================================================
\section{Statistical Methods}
\label{sec:method}
 
Our central statistical task is to estimate the correlation between two
quantities --- for example neutron star mass and the logarithm of spin period
--- from a sample of objects whose values are individually uncertain, and to do
so without the bias that naive correlation estimators suffer in the presence of
measurement error. This section sets out the model, the inference procedure,
the convergence diagnostics, and, importantly, the precise sense in which we
report statistical significance.
 
\subsection{Hierarchical bivariate model}
\label{sec:model}
For a given subpopulation we model the \emph{intrinsic} (error-free)
distribution of a pair of quantities $(M, x)$ --- where $M$ denotes the mass and
$x$ a recycling proxy such as $\log P$ --- as a bivariate Gaussian,
\begin{equation}
\begin{pmatrix} M \\ x \end{pmatrix} \sim
\mathcal{N}\!\left(
\begin{pmatrix} \mu_M \\ \mu_x \end{pmatrix},\;
\boldsymbol{\Sigma}\right), \qquad
\boldsymbol{\Sigma} = \begin{pmatrix}
\sigma_M^2 & \rho\,\sigma_M\sigma_x \\
\rho\,\sigma_M\sigma_x & \sigma_x^2
\end{pmatrix}.
\label{eq:bivar}
\end{equation}
The five parameters are the two population means $\mu_M, \mu_x$, the two
population dispersions $\sigma_M, \sigma_x$, and the correlation coefficient
$\rho \in (-1, 1)$, which is the parameter of scientific interest. A value
$\rho = 0$ corresponds to no correlation; $\rho < 0$ to an anti-correlation
(larger mass associated with smaller proxy value).
 
The reason for adopting this hierarchical formulation, rather than simply
computing a Pearson or Spearman coefficient on the measured values, is
\emph{regression dilution} (also called attenuation bias). Measurement error
scatters the observed points away from their true values, which systematically
biases a naive correlation coefficient toward zero: the more uncertain the
measurements, the weaker the apparent correlation, regardless of the true
underlying relationship. By modelling the true values as latent variables and
the measurements as noisy realisations of them, the hierarchical approach
recovers an unbiased estimate of the intrinsic correlation
\citep{Kelly2007,Mandel2019}. Concretely, for each pulsar $j$ with measured
mass $\hat M_j$ and proxy $\hat x_j$, the measured values are related to the
(unknown) true values by Gaussian measurement errors, and the likelihood is
obtained by analytically marginalising over the true values. For the mass axis
this marginalisation convolves the population variance with the measurement
variance, so that each object contributes with an effective dispersion
$\sigma_{M,\mathrm{eff},j}^2 = \sigma_M^2 + \sigma_{\hat M_j}^2$.
 
\subsection{Treatment of asymmetric uncertainties}
Pulsar mass measurements frequently have asymmetric error bars (for instance
$1.49^{+0.23}_{-0.20}\,M_\odot$), reflecting non-Gaussian posteriors from the
underlying timing analysis. Rather than symmetrising these --- which would
discard information and bias the inferred dispersion --- we adopt a two-piece
(split-normal) measurement-error model, in which the upper uncertainty
$\sigma^{+}$ is used on the high side of the mean and the lower uncertainty
$\sigma^{-}$ on the low side. This is a standard and conservative treatment of
asymmetric astronomical errors.
 
\subsection{Priors}
We adopt deliberately uninformative priors so that the data, rather than the
prior, drive the inference. The correlation coefficient $\rho$ is given a
uniform prior on $(-0.999, 0.999)$. The population means are given broad uniform
priors spanning well beyond the physical range of the data, and the dispersions
$\sigma_M, \sigma_x$ are given uniform priors on the positive half-line up to a
generous upper bound. For the mass-axis analyses we additionally bound $\mu_M$
to the physical neutron star mass range; we have verified that widening these
bounds does not affect the results.
 
\subsection{Posterior sampling and convergence}
We sample the joint posterior of the five parameters using the affine-invariant
ensemble Markov Chain Monte Carlo (MCMC) sampler \textsc{emcee}
\citep{ForemanMackey2013}, with 32--48 walkers evolved for $4\times10^4$ steps
following a burn-in of $10^4$ steps that is discarded. Convergence is assessed
by two complementary diagnostics. The Gelman--Rubin statistic $\hat R$
\citep{GelmanRubin1992} compares the variance between independent walker chains
to the variance within them; values approaching unity indicate convergence,
and we require $\hat R < 1.01$ for all reported parameters. We additionally
require an effective sample size (the number of effectively independent
posterior draws, accounting for autocorrelation within the chains) exceeding
$10^4$. All fits reported in this paper satisfy both criteria.
 
\subsection{Validation by simulation-based calibration}
To verify that our inference pipeline returns statistically correct credible
intervals, we apply simulation-based calibration \citep{Talts2018}. We generate
many synthetic datasets by drawing a true correlation $\rho_{\mathrm{true}}$
from the prior, simulating a mock population of the same size and measurement
precision as our real sample, and running the full inference. If the pipeline
is correctly calibrated, the rank of the true value within its posterior should be uniformly distributed
across the synthetic datasets. We confirm that the rank distribution is
consistent with uniform (Kolmogorov--Smirnov $p > 0.2$), demonstrating that our
credible intervals have their nominal coverage and that injected correlations
are recovered without bias.
\begin{figure}
\centering
\includegraphics[width=\columnwidth]{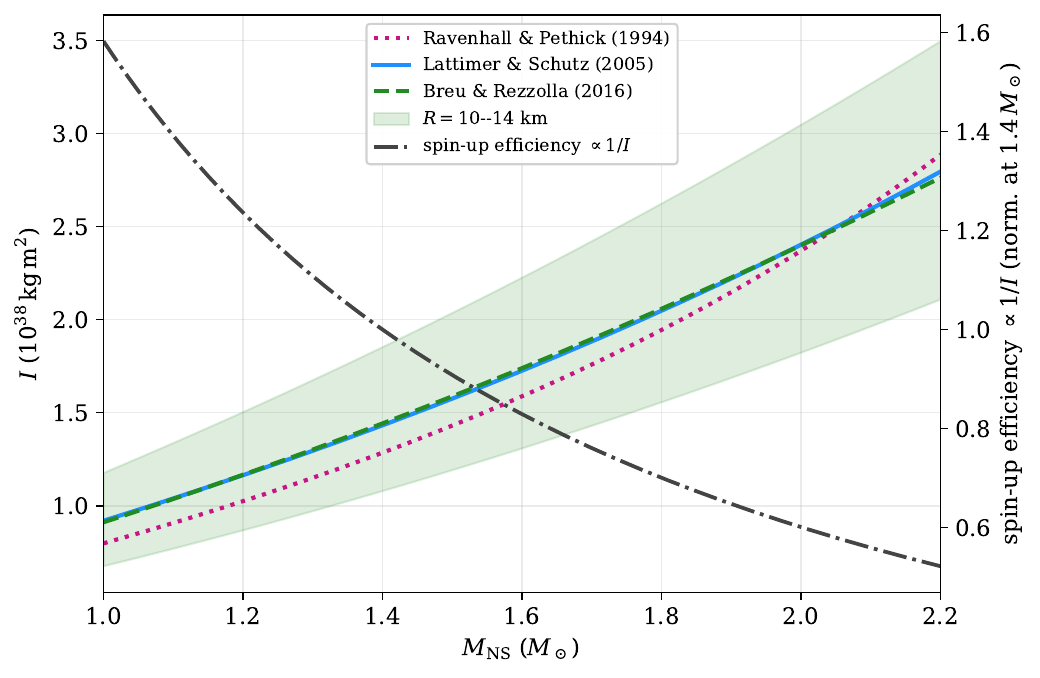}
\caption{Neutron star moment of inertia $I$ versus mass for a fixed radius
$R=12$~km, from three relations: \protect\cite{RavenhallPethick1994},
$\tilde I = 0.21/(1-2C)$; \protect\cite{LattimerSchutz2005},
$I = 0.237\,MR^2(1 + 4.2\,x + 90\,x^4)$ with $x=(M/M_\odot)/(R/{\rm km})$; and
\protect\cite{BreuRezzolla2016}, $\bar I = \sum_n a_n C^{-n}$, where $C=GM/Rc^2$
is the compactness. The shaded band spans $R=10$--$14$~km. The right-hand axis
(dot-dashed) shows the spin-up efficiency $\propto 1/I$, normalised at
$1.4\,M_\odot$, which decreases with mass --- the basis of the
moment of inertia mechanism. Because $I(M)$ is nearly linear, this mechanism is
observationally degenerate with accretion (Section~\ref{sec:massspin}).}
\label{fig:moi}
\end{figure}
\subsection{Reported quantities and the meaning of significance}
\label{sec:significance}
For each correlation we report three quantities: the posterior median of
$\rho$; its $90\%$ credible interval (the central interval containing $90\%$ of
the posterior probability); and the one-sided posterior probability
$P(\rho<0)$, which is the fraction of the posterior lying below zero. We regard
the one-sided probability as the primary summary because it is a direct,
prior-robust statement of the evidence for a correlation of a given sign.
 
We emphasise the interpretation of these quantities, because they are Bayesian
posterior probabilities and \emph{not} frequentist detection significances. A
result with $P(\rho<0) = 0.96$ means that, given the data and model, there is a
$96\%$ posterior probability that the correlation is negative; equivalently, the
$90\%$ credible interval excludes zero. In terms of a one-sided
Gaussian-equivalent, this corresponds to approximately $1.7\sigma$. By the
conventions of the field, in which ``evidence'' is conventionally associated
with $\sim 3\sigma$ and ``detection'' with $5\sigma$, our headline result
therefore constitutes \emph{moderate} or \emph{suggestive} evidence for a
mass--spin anti-correlation, not a detection. We are careful throughout to
frame it as such, and the modest sample size (Section~\ref{sec:data}) is the
principal limiting factor.
 
As a secondary, complementary summary we also compute the Savage--Dickey
density ratio \citep{Dickey1971}, which estimates the Bayes factor between the
correlated and uncorrelated ($\rho = 0$) hypotheses as the ratio of the
posterior to the prior density evaluated at $\rho = 0$. We report this for
completeness but do not lead with it, because it is sensitive to the choice of
prior width on $\rho$, whereas the one-sided probability is not.
 
\subsection{Selection effects}
\label{sec:selection-method}
The sample of binary pulsars with measured masses is shaped by selection
effects that could, in principle, imprint spurious correlations. We address
three. First, orbital motion during a survey observation Doppler-smears the
pulsed signal and degrades its detectability; we quantify this through the
orbit-averaged efficiency factor $\gamma_1$ of \citet{Bagchi2013}, which
depends on the spin and orbital periods, eccentricity, and companion mass, and
we incorporate it into a selection-corrected population likelihood through the
detection-probability normalisation of \citet{Mandel2019}. We report
correlations both with and without this correction. Second, masses measured via
the Shapiro delay require near-edge-on orbits, because at low inclination the
Shapiro signal becomes covariant with the classical R{\o}mer (Keplerian) delay
and is largely absorbed into a redefinition of the orbital parameters
\citep{Lange2001,FreireWex2010}; this renders Shapiro-derived masses an
inclination-biased subsample, an effect we test directly by correlating mass
with $\sin i$ (Section~\ref{sec:sini}). Third, wide-orbit systems suffer
negligible Doppler smearing and are the easiest to detect, so the observed
scarcity of long-orbital period systems is most plausibly astrophysical rather
than a selection artefact.
 
\subsection{Moment of inertia test}
\label{sec:moi-method}

To test whether the mass--spin correlation specifically reflects the
moment of inertia mechanism, we compare two regression models for the recycled
population. The first regresses spin period on the moment of inertia,
$\log P = \alpha + \beta\,\log_{10} I(M)$, where $I(M)$ is computed from the
equation-of-state-insensitive universal relation of \citet{BreuRezzolla2016} at
a fixed fiducial radius. The second regresses spin period directly on mass,
$\log P = \alpha + \beta\,M$. Because these two models have the same number of
parameters and the same likelihood structure, they can be compared on an equal
footing using the Bayesian Information Criterion \citep{Schwarz1978}, which
penalises model complexity; the model with the lower value is preferred.
 
% =====================================================================
\section{Results}
\label{sec:results}

\begin{figure}
\centering
\includegraphics[width=\columnwidth]{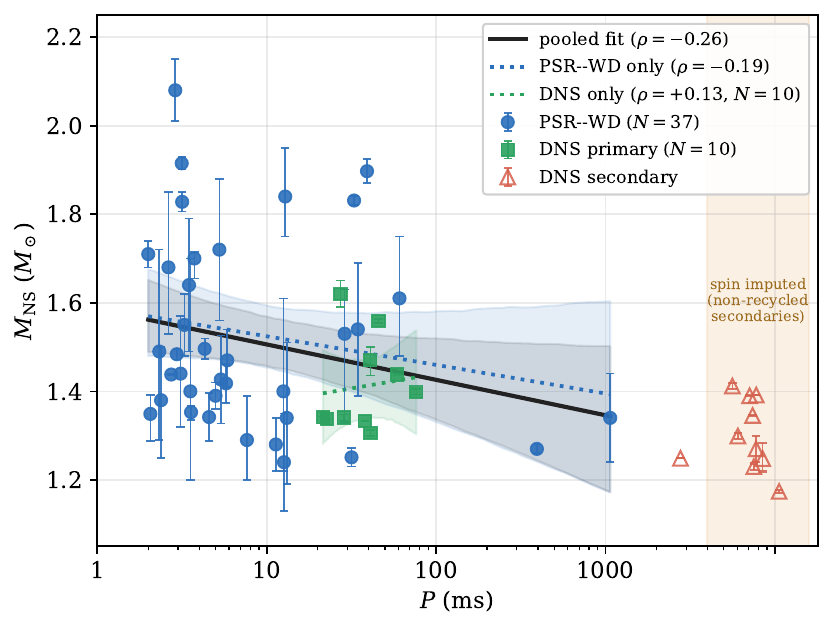}
\caption{Neutron star mass $M_{\rm NS}$ versus spin period $P$. Filled markers
are recycled primaries (PSR--WD, circles; DNS first-born, squares), on which the
correlation is measured; the pooled fit (solid) is shown with its 90\% credible
band, with the PSR--WD only and DNS only fits (dotted) illustrating the leverage
effect. Open triangles are DNS second-born neutron stars without measured spin; lacking
a spin measurement, they are placed at arbitrary long periods ($>4000$~ms, in
the shaded region above PSR~J0737$-$3039B) purely for visualisation and are
excluded from all fits. These positions are illustrative only and are not
derived from a spin-down model. PSR~J0737$-$3039B, the one second-born detected as a pulsar, appears at its true
period. Error bars are $1\sigma$ on mass.}
\label{fig:massspin}
\end{figure}

\subsection{Mass--spin correlation and the competing mechanisms}
\label{sec:massspin}
 
Our central result is moderate evidence for a mass--spin period anti-correlation
in the pooled recycled sample (all primaries, $N=47$): we infer $\rho = -0.26$,
with a $90\%$ credible interval of $[-0.47, -0.01]$ that excludes zero, and
$P(\rho<0) = 0.96$ (Table~\ref{tab:massspin}). As discussed in
Section~\ref{sec:significance}, this corresponds to roughly $1.7\sigma$ and is
suggestive rather than definitive. The result is robust to the treatment of
candidate double neutron stars: including all three cDNS strengthens it to
$\rho = -0.28$ ($P(\rho<0) = 0.97$), and including only the Galactic-disk system
J1906+0746 gives the same. It also survives the Bagchi $\gamma_1$ selection
correction, $\rho = -0.27$ ($P(\rho<0) = 0.96$), indicating that it is not an
artefact of orbital-Doppler detectability.
 
\begin{table}
\centering
\caption{Mass--spin period correlation $\rho(M,\log P)$ by sample and
candidate-DNS treatment. Column ``cDNS'' indicates whether candidate double
neutron stars are excluded (off), all included (all), or only J1906+0746
included. $^{a}$Bagchi $\gamma_1$ selection-corrected. $^{b}$Excluding the
non-recycled-pulsar systems B2303+46 and J1141$-$6545. All fits satisfy
$\hat R < 1.01$.}
\label{tab:massspin}
\begin{tabular}{llccc}
\hline
Sample & cDNS & $N$ & $\rho$ (90\% CI) & $P(\rho<0)$ \\
\hline
All primaries & off    & 47 & $-0.26\,[-0.47,-0.01]$ & 0.96 \\
All primaries & all    & 50 & $-0.28\,[-0.49,-0.04]$ & 0.97 \\
All primaries & J1906  & 48 & $-0.28\,[-0.49,-0.04]$ & 0.97 \\
All primaries$^{a}$ & off & 47 & $-0.27\,[-0.50,-0.01]$ & 0.96 \\
All primaries$^{b}$ & off & 45 & $-0.18\,[-0.41,+0.09]$ & 0.87 \\
PSR--WD only  & ---    & 37 & $-0.19\,[-0.45,+0.10]$ & 0.86 \\
DNS primary   & off    & 10 & $+0.13\,[-0.42,+0.61]$ & 0.36 \\
\hline
\end{tabular}
\end{table}
 
\begin{figure}
\centering
\includegraphics[width=\columnwidth]{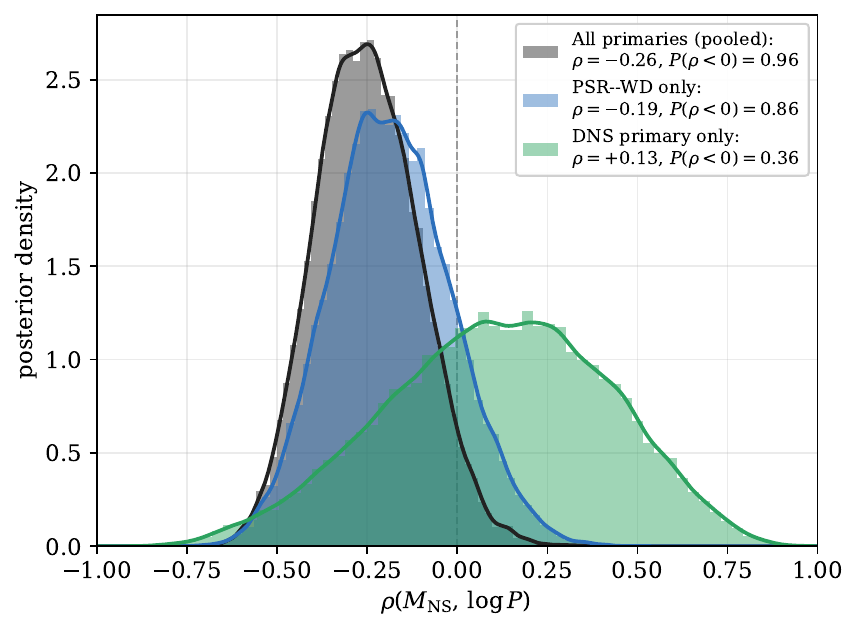}
\caption{Posterior distributions of the mass--spin correlation
$\rho(M_{\rm NS},\log P)$ for the pooled sample (black) and the PSR--WD only and
DNS only subpopulations. The pooled posterior places 96\% of its probability at
$\rho<0$ ($\sim1.7\sigma$). Colours and values match
Figure~\ref{fig:massspin}.}
\label{fig:rhopost}
\end{figure}

Two features of Table~\ref{tab:massspin} warrant comment. First, the PSR--WD
systems alone yield only a suggestive trend ($\rho = -0.19$, with the $90\%$
interval crossing zero); the pooled significance derives substantially from the
inclusion of the precisely measured DNS primaries. This is a statistical
leverage effect rather than an internal trend within the DNS sample, which is
individually unconstrained ($\rho = +0.13$ at $N=10$, with 
the posterior
is broad and only weakly informative, with its median shifted towards positive
$\rho$). The three posteriors are shown in Figure~\ref{fig:rhopost}. Intriguingly, the DNS subpopulation alone leans to the opposite, positive side
($\rho=+0.13$, $N=10$) --- the direction expected from the moment of inertia
mechanism rather than from accretion --- though with only ten systems this is
not statistically significant and may simply reflect that these
birth-mass-dominated systems do not share the accretion-driven trend of the
PSR--WD population. As Figure~\ref{fig:massspin} illustrates, the DNS cloud
occupies a region of longer spin period and lower mass than the PSR--WD cloud;
pooling the two extends the negative diagonal in the mass--spin plane and
steepens the inferred slope. We emphasise that a genuine same-sign trend is
present within the PSR--WD sample, so that pooling amplifies a real correlation
rather than manufacturing one, but we disclose the leverage effect explicitly as
it bears on the interpretation. Second, excluding the two systems whose pulsars
are non-recycled despite having white dwarf companions (B2303+46 and
J1141$-$6545, identifiable by their long spin periods and reversed formation
order) weakens the correlation to $\rho = -0.18$, indicating that these systems
contribute appreciably to the signal; we retain them in the primary analysis as
bona fide members of the mass--spin plane while flagging their influence.
The negative sign of the correlation is consistent with accretion-driven
recycling. To ask whether it specifically reflects the moment of inertia
mechanism, we compare the regression of spin period on $\log I(M)$ (model B1,
inferred slope $\beta = -2.05$) with the regression on mass directly (model B2,
$\beta = -0.78$). The two models are statistically indistinguishable, with a
Bayesian Information Criterion difference of only $\Delta\mathrm{BIC} = -0.2$,
because $\log I(M)$ is very nearly linear in $M$ over the narrow mass range
spanned by neutron stars. We therefore detect a mass--spin trend consistent
with recycling, but we cannot isolate the moment of inertia mechanism from
generic accretion with the present sample (Figure~\ref{fig:moi}). 
We return to how this degeneracy might be broken observationally in Section~\ref{sec:forecast}.

\subsection{Mass versus other proxies}
\label{sec:proxies}
\begin{figure}
\centering
\includegraphics[width=\columnwidth]{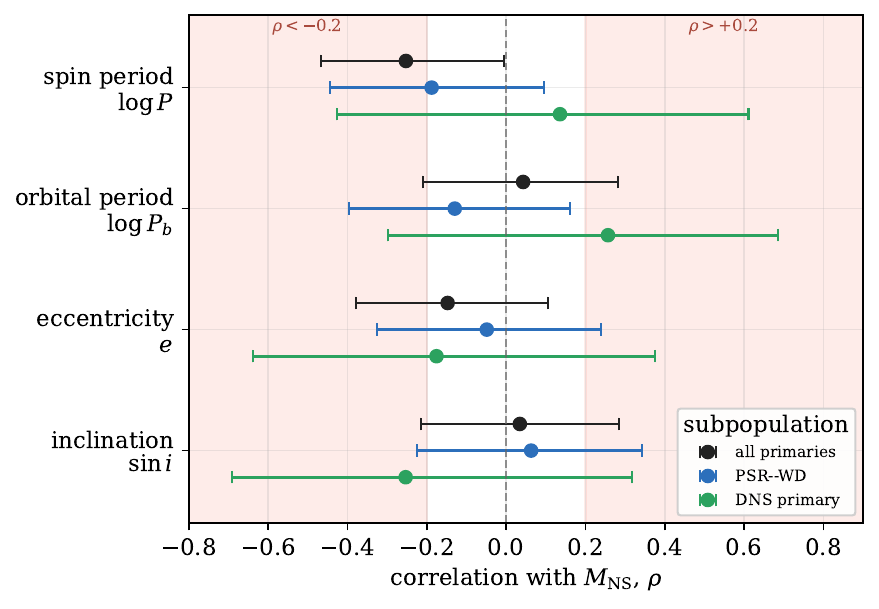}
\caption{Correlation of neutron star mass $M_{\rm NS}$ with each recycling proxy,
for the pooled sample (black), PSR--WD only (blue), and DNS primaries (green).
Shaded regions mark $|\rho|>0.2$ for visual reference. Only the spin period shows an
appreciable correlation; the orbital period and inclination are consistent with
zero, while the eccentricity shows at most a weak trend that is not
statistically significant.}
\label{fig:proxies}
\end{figure}

Mass correlates significantly only with spin period among the recycling proxies
we examine (Table~\ref{tab:proxies}, Figure~\ref{fig:proxies}). The correlation
of mass with orbital period is consistent with zero ($\rho = +0.04$); with
eccentricity it is weak and not significant ($\rho = -0.15$, $P(\rho<0)=0.83$);
and with orbital inclination it is consistent with zero ($\rho = +0.02$). The
spin period is thus the only proxy that carries a significant imprint of the
recycling process onto the neutron star mass.
 
\begin{table}
\centering
\caption{Correlation of pulsar mass with each recycling proxy in the pooled
recycled sample (all primaries, $N=47$; the $\sin i$ test uses $N=46$ owing to
one system with an undefined inclination).}
\label{tab:proxies}
\begin{tabular}{lccc}
\hline
Proxy & $\rho$ (90\% CI) & $P(\rho<0)$ & Verdict \\
\hline
Spin period $\log P$      & $-0.26\,[-0.47,-0.01]$ & 0.96 & significant \\
Orbital period $\log P_b$ & $+0.04\,[-0.21,+0.29]$ & 0.40 & null \\
Eccentricity $e$          & $-0.15\,[-0.38,+0.11]$ & 0.83 & weak \\
$\sin i$                  & $+0.02\,[-0.24,+0.27]$ & 0.46 & null \\
\hline
\end{tabular}
\end{table}
\subsection{Inclination selection check}
\label{sec:sini}
The null correlation between mass and $\sin i$ holds across all subpopulations
(pooled $\rho = +0.02$) and is robust to the treatment of candidate double
neutron stars. We note that the sample-mean inclination is high,
$\langle\sin i\rangle \approx 0.90$, compared with the value $\approx 0.79$
expected for randomly oriented orbits; this confirms the anticipated
over-representation of near-edge-on systems among those with Shapiro-delay mass
measurements (Section~\ref{sec:selection-method}). Crucially, however, this
inclination selection is not mass-dependent --- mass and $\sin i$ are
uncorrelated --- and therefore does not bias the mass--spin result.
\begin{figure}
\centering
\includegraphics[width=0.5\textwidth]{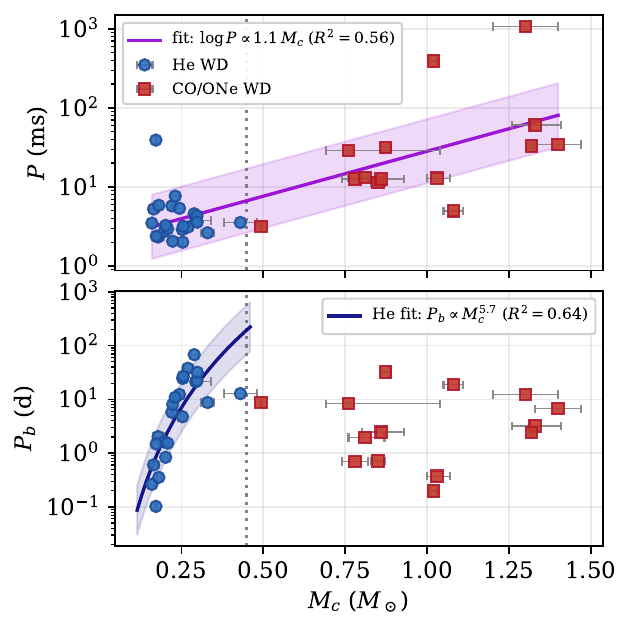}
\caption{Recycled-pulsar spin period $P$ (top) and orbital period $P_b$
(bottom) versus white dwarf companion mass $M_c$, sharing the $M_c$ axis.
Helium (circles, blue edge) and CO/ONe (squares, red edge) companions are
divided at $M_c = 0.45\,M_\odot$ (dotted line). Spin period increases with
companion mass (top; log-linear fit $\log_{10} P \propto 1.1\,M_c$, $R^2=0.56$).
The helium systems follow a steep orbital period--companion mass relation
(bottom; $P_b \propto M_c^{5.7}$, $R^2=0.64$), consistent in slope with the
helium white dwarf relation of \protect\citet{TaurisSavonije1999}, whereas the
CO/ONe systems show no such trend ($R^2\simeq0$), confirming that the relation
is specific to the helium white dwarf channel. Shaded bands show the $1\sigma$
scatter about each fit; horizontal bars are $1\sigma$ uncertainties on $M_c$.}
\label{fig:companion}
\end{figure} 
\subsection{Companion type and mass}
\label{sec:companion}

Within the PSR--WD systems, the recycled pulsar spin period correlates strongly
with companion mass (Spearman $\rho = +0.59$; Figure~\ref{fig:companion}). This correlation is driven by the
distinction between helium and CO/ONe white dwarf companions rather than by a
continuous trend within either class. Helium white dwarfs follow the expected
orbital period--companion mass relation strongly (Spearman $\rho = +0.85$; lower panel of Figure~\ref{fig:companion}),
validating our mass-based companion classification, whereas the CO/ONe systems
show no such trend (Spearman $\rho = +0.10$).

\begin{figure}
\centering
\includegraphics[width=0.5\textwidth]{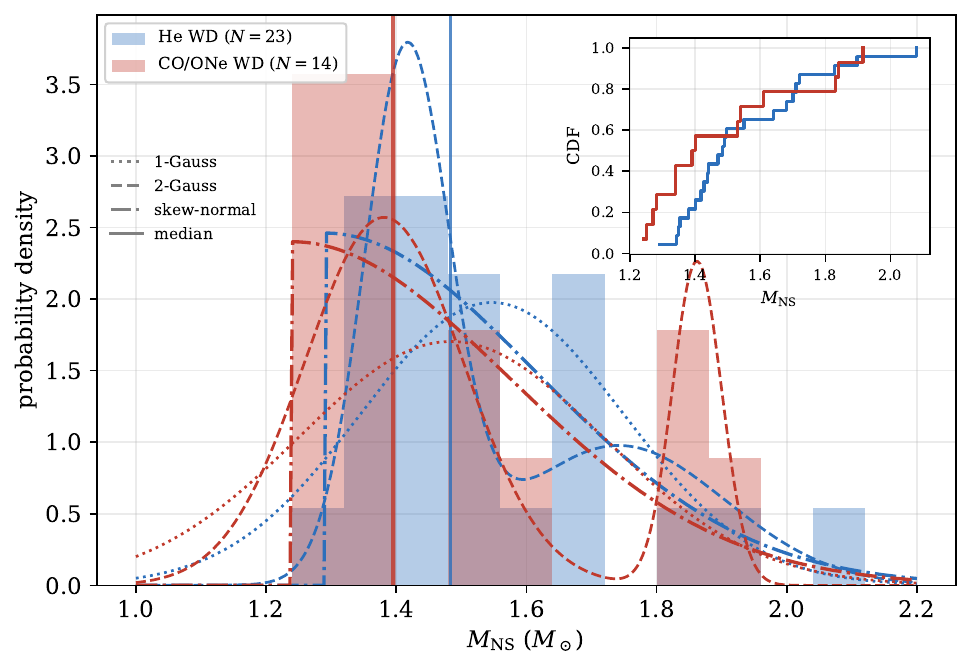}
\caption{Neutron star mass distributions for helium versus CO/ONe white dwarf
companions, divided at $M_c = 0.45\,M_\odot$ \citep{DCruz1996}. Curves show
single-Gaussian (dotted), two-Gaussian (dashed), and skew-normal (dash-dotted)
fits; solid vertical lines mark the medians and the inset shows the cumulative
distributions. Neutron stars with helium companions are marginally more massive
(means $1.547$ versus $1.484\,M_\odot$; medians $1.484$ versus
$1.395\,M_\odot$), but the difference is not statistically significant
(KS $p = 0.11$). That the median lies below the mean in both samples indicates a
common right-skewed shape with an extended high-mass tail.}
\label{fig:wdtype}
\end{figure}

We further find that neutron stars with helium white dwarf companions are
marginally more massive than those with CO/ONe companions,  
with mean masses of $1.547$ and $1.484\,M_\odot$ respectively, a difference of
$+0.06\,M_\odot$
(Figure~\ref{fig:wdtype}). This offset is not statistically significant
(two-sample Kolmogorov--Smirnov $p = 0.11$) and is sensitive to the placement
of the He/CO boundary, rising to $\simeq0.11\,M_\odot$ for a division at
$0.50\,M_\odot$ but remaining below $2\sigma$ for any reasonable choice; the
Bayesian Information Criterion favours a common distribution. The two
populations share a common right-skewed shape and differ principally in
location. The direction of the offset is that expected if helium-white dwarf
systems, which descend from longer and more stable episodes of mass transfer,
accrete more mass onto the neutron star --- the same accretion signature seen
in the mass--spin correlation, here resolved by companion type. Detailed binary-evolution models find that the helium-star mass-transfer phase
dominates the recycling of the neutron star, with closer orbits permitting more
extensive accretion \citep{Chattaraj2025}; while these models concern double
neutron star rather than pulsar--white dwarf formation, the same qualitative
link between a more extended helium-donor mass-transfer phase and greater
neutron star accretion supports the direction of the offset we observe. We
caution that the marginal significance of our result does not yet require it.

\subsection{Mass distributions across populations}
\label{sec:massdist}
\begin{figure*}
\centering
\includegraphics[width=\textwidth]{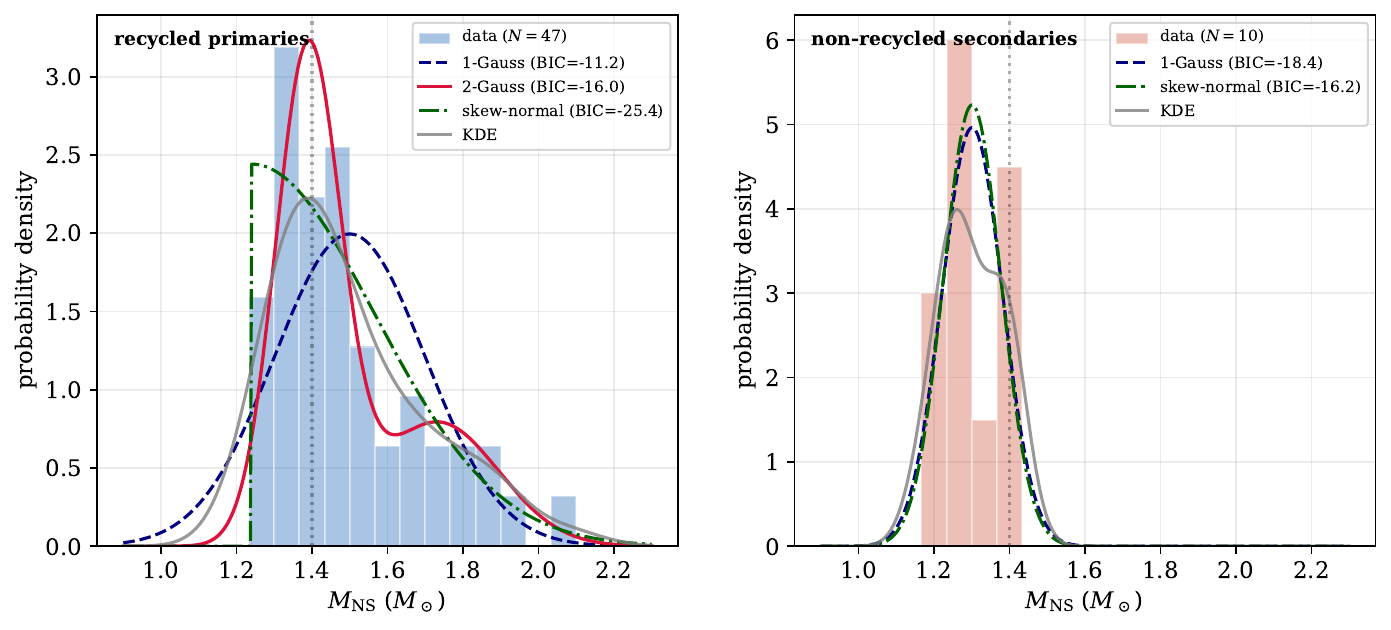}
\caption{Mass distributions of the recycled primaries ($N=47$, left) and
non-recycled secondaries ($N=10$, right), each with single-Gaussian,
skew-normal, and kernel-density fits; a two-Gaussian mixture is also shown for
the recycled sample, where the sample is large enough to constrain it. For the
recycled primaries the two-Gaussian mixture (BIC $=-16.0$) is preferred over a
single symmetric Gaussian (BIC $=-11.2$), but a single right-skewed component
provides the best description overall (BIC $=-25.4$), indicating that the
apparent bimodality is more economically explained by a single population with
a high-mass tail than by two distinct components.}
\label{fig:massdist}
\end{figure*}

Turning from correlations to the distributions themselves, we find that the
recycled primaries are more massive than the non-recycled DNS secondaries by
$+0.20\,M_\odot$ ($1.50$ versus $1.30\,M_\odot$; two-sample Kolmogorov--Smirnov
$p = 0.004$, with $\Delta\mathrm{BIC} = -10.8$ favouring distinct
distributions), consistent with the offset reported by \citet{Kiziltan2013}.
Within the recycled class, the PSR--WD and DNS-primary masses differ only
marginally ($+0.11\,M_\odot$). Comparing single-Gaussian, two-Gaussian, skew-normal, and kernel-density fits
to each subpopulation (Figure~\ref{fig:massdist}), we find for the recycled
primaries that a two-Gaussian mixture ($\mathrm{BIC}=-16.0$) is favoured over a
single symmetric Gaussian ($\mathrm{BIC}=-11.2$), consistent with the
bimodality reported in earlier studies \citep{Antoniadis2016,Kiziltan2013}, but
that a single right-skewed component yields the lowest BIC ($-25.4$),
describing the same structure with fewer parameters. The apparent bimodality is
therefore more parsimoniously characterised as a skewed population with a
high-mass tail than as two distinct peaks. This is consistent with recent findings that the
recycled and Gaia astrometric neutron star populations share a narrow peak near
$1.3\,M_\odot$ with a broad high-mass tail \citep{Su2025}, and that
accretion-corrected birth masses favour a unimodal turn-on distribution over a
double Gaussian \citep{You2025}. The non-recycled DNS secondaries form a single
narrow peak near $1.30\,M_\odot$.
\subsection{Companion mass and eccentricity in double neutron stars}
\label{sec:ecc}
For the double neutron star systems, the companion (second-born) mass correlates
strongly with orbital eccentricity, $\rho(M_c, e) = +0.82$ (with a $90\%$
credible interval of $[+0.48, +0.94]$ and $P(\rho>0) = 1.00$, for $N=10$). This
confirms, within a hierarchical framework, a trend that has previously been
noted qualitatively \citep{Tauris2017,AndrewsMandel2019,Sengar2022}: low-kick
electron-capture or ultra-stripped supernovae are predicted to produce both low
orbital eccentricities and low second-born neutron star masses. We caution that
this result rests on only ten systems and should be regarded as indicative
rather than definitive. For completeness, the companion mass--$\sin i$
correlation is weak and consistent with zero given the small sample
($\rho = -0.37$, $N=10$). We also note that the apparent anti-correlation
between pulsar and companion mass ($\rho = -0.28$) is at least partly a
consequence of the constraint on the total mass imposed by the
periastron-advance measurement, and we do not interpret it further.
 
% =====================================================================
\section{Discussion}
\label{sec:discussion}
 
\subsection{A two-channel picture}
Our results are coherently explained if the PSR--WD and DNS systems represent
two distinct regimes of neutron star mass assembly. The PSR--WD systems descend
from low-mass X-ray binaries, in which mass transfer proceeds over extended
timescales and can substantially increase the neutron star mass. Their neutron
stars are therefore accretion-grown, and the signatures of this accretion are
evident in our data: the mass--spin anti-correlation, the marginal He/CO-ONe
mass offset, and the spin--companion mass relation are all manifestations of the
same underlying process. The double neutron stars, by contrast, accrete very
little --- at most $\sim 0.02\,M_\odot$ during the brief recycling episode
\citep{Tauris2017} --- so their masses encode birth conditions rather than
recycling history. In this picture the companion mass--eccentricity correlation
in the DNS systems is a signature of the second supernova and its kick, not of
accretion. The mass--spin correlation thus originates physically in the
accretion channel, while the contribution of the DNS primaries to its
statistical significance reflects the \emph{location} of the
birth-mass-dominated DNS systems in the mass--spin plane rather than an
accretion process operating within them.
\begin{figure}
\centering
\includegraphics[width=\columnwidth]{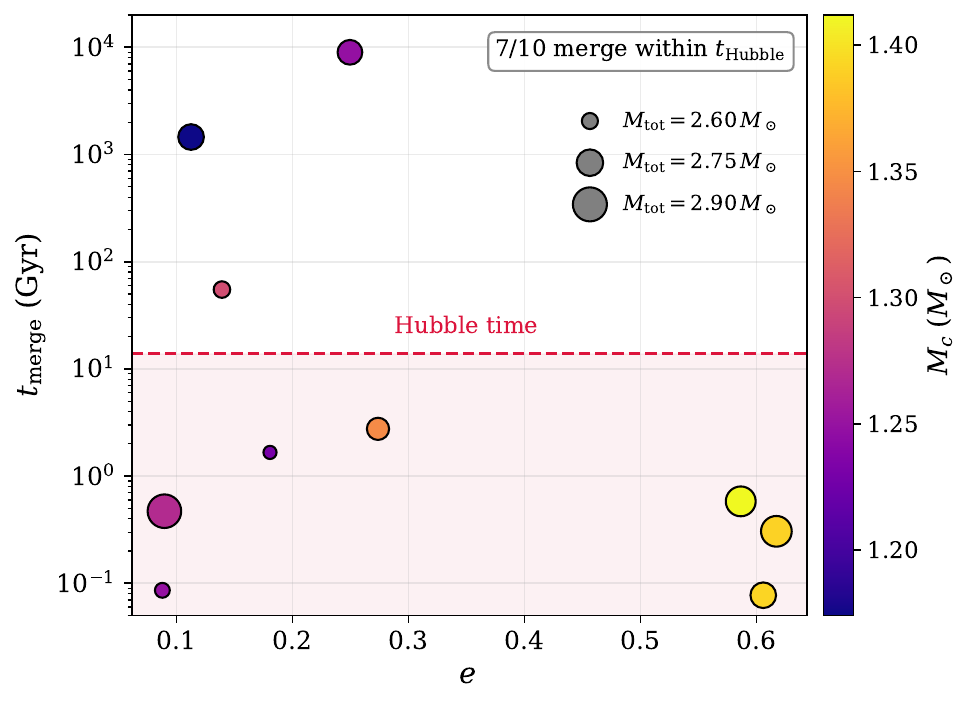}
\caption{gravitational wave merger time, computed from the Peters (1964) formula,
versus eccentricity for the ten double neutron star systems. Point colour encodes
companion mass $M_c$ and symbol size the total mass $M_{\rm tot}$. Seven of ten
merge within a Hubble time (shaded); the fast-merging systems are preferentially
eccentric and of higher companion mass.}
\label{fig:peters}
\end{figure}
\subsection{The origin of the high-mass component}
Our findings bear on the unresolved question of whether the high-mass neutron
stars in the recycled distribution are born massive or grown by accretion
(Section~\ref{sec:intro}). Producing a $\sim 2\,M_\odot$ neutron star by
accretion requires the transfer of several tenths of a solar mass, as inferred
for the most massive recycled objects \citep{Romani2022}, whereas supernova
models struggle to produce enough massive neutron stars at birth
\citep{Antoniadis2016}. Two aspects of our analysis favour an accretion
contribution: the high-mass component appears in the accretion channel
(PSR--WD) and not in the DNS systems, and the mass--spin correlation
independently links higher mass to more rapid recycling. The preference for a
skew-normal over a two-Gaussian description (Section~\ref{sec:massdist}) is
moreover consistent with the emerging view \citep{You2025,Su2025} that the
recycled distribution is better understood as a skewed population with a
high-mass tail than as two cleanly separated peaks.
 
\subsection{Connection to the gravitational wave population}
Both of our correlations connect to the population of compact binaries observed
in gravitational waves. First, the companion mass--eccentricity correlation has
a direct consequence for which double neutron stars merge. Because eccentric
binaries inspiral more rapidly --- in the limiting form
$t_{\mathrm{merge}} \propto (1-e^2)^{7/2}$ at fixed semi-major axis
\citep{Peters1964} --- the gravitational-wave-merging subset of the DNS
population is preferentially eccentric, and, through the positive
companion mass--eccentricity correlation, mildly enriched in heavier
second-born neutron stars. We illustrate this with the Galactic DNS sample:
seven of the ten systems merge within a Hubble time, and the merging subset has
roughly twice the mean eccentricity of the non-merging subset ($0.35$ versus
$0.17$) and a slightly higher mean companion mass ($1.33$ versus
$1.24\,M_\odot$; Figure~\ref{fig:peters}). This is a mild and testable bias
rather than a large effect. Second, the mass--spin correlation implies that the
recycled neutron star in a merging double neutron star will have spun down below
radio detectability long before merger, so that the radio and gravitational-wave
populations sample the same recycled systems at different evolutionary epochs.
The component masses of GW170817 are consistent with the Galactic radio
distribution \citep{Abbott2017}, validating our mass scale, whereas the more
massive GW190425 \citep{Abbott2020} may indicate a formation channel
under-represented in the local radio sample \citep{NairStevenson2025}  ---
although its outlier significance relaxes from $\sim 5\sigma$ to $\sim 3\sigma$
when uncertainties in the Galactic-population model are propagated
\citep{Farrow2019}. More broadly, the gravitational-wave-detected neutron star
population is consistent with a flatter mass distribution extending to higher
masses than the Galactic radio population \citep{Landry2021}. A quantitative
forward model linking our radio-inferred correlations to the
gravitational wave population is the subject beyond the scope of this work, and may be explored in later project(s).
\begin{figure}
\centering
\includegraphics[width=\columnwidth]{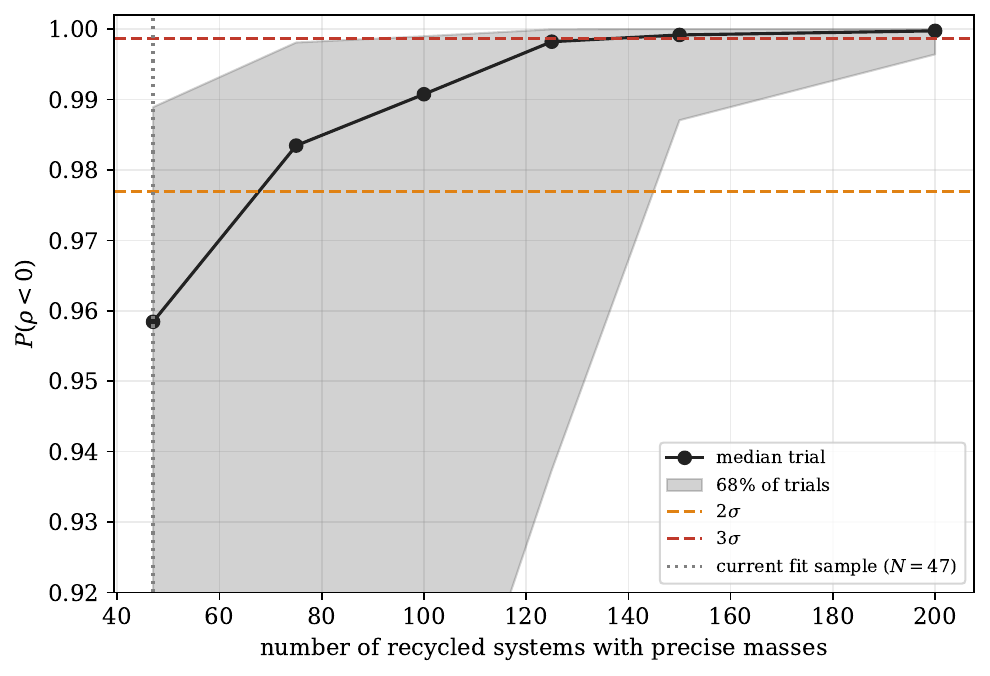}
\caption{Forecast significance $P(\rho<0)$ of the mass--spin correlation as a
function of sample size, taking the inferred parameters as ground truth (median
and 68\% range over mock catalogues). The dotted line marks the current fit
sample ($N=47$). A $2\sigma$ result is expected near $N\simeq100$, while a
confident $3\sigma$ measurement requires $N\gtrsim125$--$150$.}
\label{fig:forecast}
\end{figure}
\subsection{Prospects: sample size required for a confident detection}
\label{sec:forecast}
The moderate significance of our headline correlation
(Section~\ref{sec:significance}) raises the practical question of how much the
sample must grow before the mass--spin anti-correlation can be confirmed at the
level conventionally regarded as evidence. To address this we perform a forecast:
adopting our inferred correlation and population parameters as ground truth
($\rho=-0.26$, $\mu_M=1.50\,M_\odot$, $\sigma_M=0.20\,M_\odot$,
$\sigma_{\log P}=0.62$, and the measured mass-error scale), we simulate mock
samples of increasing size, propagate realistic measurement uncertainties, and
re-run the full hierarchical inference on many independent realisations at each
sample size, recording the distribution of $P(\rho<0)$
(Figure~\ref{fig:forecast}).

The forecast indicates that the correlation is on a credible path to
confirmation, but that this requires a substantial increase in the number of
recycled systems with precise mass measurements. At the current sample size
($N\simeq47$) the median realisation yields $P(\rho<0)\simeq0.95$, consistent
with our measured value of $0.96$; only $\sim7\%$ of realisations would reach
the $3\sigma$ level by chance. Doubling the sample to $N\simeq100$ raises the
median to $P(\rho<0)\simeq0.99$ ($\sim2\sigma$), with
roughly three quarters of realisations exceeding $2\sigma$. A robust
$3\sigma$-equivalent result is reached by the median realisation only for
$N\gtrsim125$, and even at $N=200$ approximately one third of realisations fall
short of $3\sigma$. This flattening is intrinsic to a correlation of moderate
strength: a true $|\rho|\simeq0.26$ cannot be driven to high significance by
modest increases in sample size alone. We therefore expect the correlation to
firm up to $\sim2\sigma$ as the precise-mass sample approximately doubles --- plausibly on a decadal timescale given ongoing and forthcoming pulsar surveys
with FAST, MeerKAT, and ultimately the Square Kilometre Array --- while a
confident $3\sigma$ measurement will likely require the SKA-era sample. 
We caution, however, that discovering a recycled binary is a necessary but not
sufficient condition for entry into our sample: a precise mass requires the
measurement of post-Keplerian parameters, typically over years of dedicated
timing. While FAST has already discovered $\sim\mathcal{O}(10^2)$ binary pulsars
through its Galactic Plane Pulsar Snapshot survey \citep{Han2025}, and MeerKAT
continues to deliver Shapiro-delay masses, the conversion of these discoveries
into precise component masses is comparatively slow. Doubling the precise-mass
sample of recycled binaries is therefore more realistically an SKA-era
achievement than a near-term one, consistent with forecasts of a factor
$\sim5$ increase in high-timing-precision sources with next-generation
facilities. We note
the corollary that, should the true correlation be weaker than inferred here,
even a doubled sample may remain inconclusive; this reinforces our framing of
the present result as suggestive evidence rather than a detection.

A complementary route to disentangling the accretion and moment of inertia
mechanisms does not require a larger sample, but rather a different kind of
measurement. Because both mechanisms are inferred here only from recycled
neutron stars, whose spin has been reset by accretion, the two remain
degenerate in our data (Section~\ref{sec:massspin}). A \emph{non-recycled}
neutron star with both a precisely measured mass and a measured spin period
would break this degeneracy directly, since its spin reflects its birth and
subsequent evolution without the confounding effect of recycling. The
second-born neutron star in the double pulsar, PSR~J0737$-$3039B, is at present
the only such object with both quantities measured; its slow spin and ordinary
mass are consistent with either scenario but cannot by themselves distinguish
them. The discovery and timing of further double neutron star systems in which
the second-born star is itself detectable as a pulsar --- a natural target for
ongoing and forthcoming surveys with MeerKAT, CHIME, and ultimately the SKA ---
would provide the cleanest empirical test of whether the neutron star moment of
inertia shapes the recycled mass distribution.

\subsection{Caveats}
Several limitations bound our conclusions. The sample is modest ($N \sim 50$)
and the individual subpopulations are underpowered; the DNS only correlations,
including the companion mass--eccentricity result, rest on $N=10$ and are
indicative rather than definitive. The threshold of approximately twenty precise
double neutron star component masses identified by \citet{Farrow2019} marks the
point at which several of these questions become decisively testable. The
moment of inertia and accretion mechanisms cannot be distinguished here owing to
the near-linearity of $I(M)$ over the neutron star mass range. The statistical
significance of the headline correlation depends partly on the location of the
DNS systems in the mass--spin plane (Section~\ref{sec:massspin}). The He/CO-ONe
classification is imperfect near its metallicity-dependent boundary, and the
associated mass offset is of marginal significance. Finally, the $\gamma_1$
selection correction is a first-order treatment rather than a full
injection-and-recovery analysis.

% =====================================================================
\section*{Acknowledgements}
DC acknowledges the Gordon and Betty Moore Foundation for funding this research through Grant GBMF12341. DC thanks Vivek Venkatraman Krishnan, Simon Stevenson and Marcus Lower for their insightful comments. 

%%%%%%%%%%%%%%%%%%%%%%%%%%%%%%%%%%%%%%%%%%%%%%%%%%
\section*{Data Availability}
The data utilised for this work will be freely available upon reasonable request to the corresponding author
 
% The inclusion of a Data Availability Statement is a requirement for articles published in MNRAS. Data Availability Statements provide a standardised format for readers to understand the availability of data underlying the research results described in the article. The statement may refer to original data generated in the course of the study or to third-party data analysed in the article. The statement should describe and provide means of access, where possible, by linking to the data or providing the required accession numbers for the relevant databases or DOIs.

% =====================================================================
\appendix
\section{The Data Sample}
\label{sec:appendix-data}
Table~\ref{tab:data} lists the full sample used in this analysis: spin period
$P$, orbital period $P_b$, eccentricity $e$, pulsar mass $M_p$, companion mass
$M_c$, system type, and (for PSR--WD systems) the white dwarf type inferred from
the companion mass.

\begin{table*}
\centering
\caption{The binary pulsar sample. Spin period $P$ in ms, orbital period $P_b$
in days, and eccentricity $e$. Orbital periods are measured to a fractional
precision of $\lesssim10^{-6}$ and eccentricities to better than the displayed
precision, so their (negligible) uncertainties are omitted. Masses of the neutron star $M_{\rm NS}$ and companion
$M_c$ in $M_\odot$ carry $1\sigma$ uncertainties in parentheses, expressed on
the final quoted digits: e.g.\ $1.338(14)$ means $1.338\pm0.014$, and
$1.490(+23)(-20)$ denotes an asymmetric uncertainty $^{+0.023}_{-0.020}$. WD
type is He or CO/ONe for pulsar--white dwarf systems (Section~\ref{sec:data});
ellipses denote quantities not applicable or not measured. Masses are taken from
the compilation of pulsar mass measurements maintained by
P.~C.~C.~Freire\protect\footnote{\url{https://www3.mpifr-bonn.mpg.de/staff/pfreire/NS_masses.html},
accessed 2026 June.} and the original timing references therein (final column),
and cross-checked against the ATNF Pulsar Catalogue \protect\citep{Manchester2005}.
Systems with $M_c$ within $0.05\,M_\odot$ of the $0.45\,M_\odot$ He/CO division
are flagged ``boundary/X'', where X is the class assigned by the
$M_c = 0.45\,M_\odot$ cut.}
\label{tab:data}
\begin{tabular}{lccccccl}
\hline
Pulsar & $P$ (ms) & $P_b$ (d) & $e$ & $M_{\rm NS}$ ($M_\odot$) & $M_c$ ($M_\odot$) & WD type & Reference \\
\hline
\multicolumn{8}{l}{\textit{Pulsar--white dwarf systems}}\\
\hline
J1802$-$2124 & 12.65 & 0.6989 & 2.48e-06 & 1.240(110) & 0.780(40) & CO & \citealt{Ferdman2010} \\
J2045+3633 & 31.68 & 32.3 & 0.01721 & 1.251(21) & 0.873(15) & CO & \citealt{McKee2020} \\
J1141$-$6545 & 393.86 & 0.1976 & 0.1719 & 1.270(10) & 1.020(10) & CO & \citealt{Bhat2008} \\
J2023+2853 & 11.33 & 0.7182 & 1.31e-05 & 1.280(60) & 0.850(20) & CO & \citealt{Yang2025RAA} \\
J1918$-$0642 & 7.65 & 10.91 & 2.034e-05 & 1.290(95) & 0.231(10) & He & \citealt{Arzoumanian2018} \\
J1949+3106 & 13.14 & 1.95 & 4.312e-05 & 1.340(160) & 0.810(+60)(-50) & CO & \citealt{Zhu2019} \\
B2303+46 & 1067.24 & 12.34 & 0.6584 & 1.340(100) & 1.300(100) & CO & \citealt{Thorsett1999b} \\
J1713+0747 & 4.57 & 67.83 & 7.494e-05 & 1.342(54) & 0.289(+13)(-11) & He & \citealt{Arzoumanian2018} \\
J1543$-$5149 & 2.06 & 8.061 & 2.133e-05 & 1.349(+43)(-61) & 0.223(+5)(-6) & He & \citealt{Colom2025} \\
J2234+0611 & 3.58 & 32 & 0.1293 & 1.353(+14)(-17) & 0.298(+15)(-12) & He & \citealt{Stovall2019} \\
J2043+1711 & 2.38 & 1.482 & 4.89e-06 & 1.380(125) & 0.173(10) & He & \citealt{Arzoumanian2018} \\
J0514$-$4002A & 4.99 & 18.79 & 0.888 & 1.390(30) & 1.080(30) & CO & \citealt{Dixon2025} \\
J2053+4650 & 12.55 & 2.453 & 8.9e-06 & 1.400(195) & 0.860(65) & CO & \citealt{Berthereau2023} \\
J1933$-$6211 & 3.54 & 12.82 & 1.26e-06 & 1.400(+300)(-200) & 0.430(50) & boundary/He & \citealt{Geyer2023} \\
J0437$-$4715 & 5.76 & 5.741 & 1.918e-05 & 1.418(44) & 0.221(4) & He & \citealt{Reardon2024} \\
B1855+09 & 5.36 & 12.33 & 2.17e-05 & 1.427(99) & 0.244(13) & He & \citealt{Arzoumanian2018} \\
J0337+1715 & 2.73 & 1.629 & 0.0006989 & 1.438(1) & 0.198(15) & He & \citealt{Voisin2020} \\
J1012$-$4235 & 3.10 & 37.97 & 0.0003457 & 1.440(125) & 0.270(16) & He & \citealt{Gautam2024} \\
J1738+0333 & 5.85 & 0.3548 & 3.4e-07 & 1.470(65) & 0.181(+7)(-5) & He & \citealt{Antoniadis2012} \\
J1909$-$3744 & 2.95 & 1.533 & 1.15e-07 & 1.484(12) & 0.208(2) & He & \citealt{Arzoumanian2018} \\
J0218+4232 & 2.32 & 2.029 & 6.3e-06 & 1.490(215) & 0.179(17) & He & \citealt{Tang2024} \\
J1950+2414 & 4.30 & 22.19 & 0.07981 & 1.496(23) & 0.297(+5)(-4) & He & \citealt{Zhu2019} \\
J0621+1002 & 28.85 & 8.319 & 0.002457 & 1.530(+100)(-200) & 0.760(+280)(-70) & CO & \citealt{Kasian2012} \\
J1227$-$6208 & 34.53 & 6.721 & 0.001149 & 1.540(150) & 1.400(70) & CO & \citealt{Colom2024} \\
J1910$-$5958A & 3.27 & 0.8371 & 8.2e-07 & 1.550(70) & 0.202(6) & He & \citealt{Corongiu2023} \\
J1528$-$3146 & 60.82 & 3.18 & 0.0002137 & 1.610(135) & 1.330(75) & CO & \citealt{Berthereau2023b} \\
J0751+1807 & 3.48 & 0.2631 & 5e-07 & 1.640(150) & 0.160(10) & He & \citealt{Desvignes2016} \\
J1125$-$6014 & 2.63 & 8.753 & 6.15e-07 & 1.680(160) & 0.330(20) & He & \citealt{Sengar2023} \\
J1748$-$2446ap & 3.74 & 21.39 & 0.9052 & 1.700(+15)(-45) & 0.294(+46)(-15) & He & \citealt{Prager2024} \\
J0955$-$6150 & 2.00 & 24.58 & 0.1175 & 1.710(30) & 0.254(2) & He & \citealt{Serylak2022} \\
J1012+5307 & 5.26 & 0.6047 & 1.2e-06 & 1.720(160) & 0.165(15) & He & \citealt{Mata2020} \\
J1946+3417 & 3.17 & 27.02 & 0.1345 & 1.828(22) & 0.256(2) & He & \citealt{Barr2017} \\
J2222$-$0137 & 32.82 & 2.446 & 0.0003809 & 1.831(10) & 1.319(4) & CO & \citealt{Guo2021} \\
J1943+2210 & 12.87 & 0.372 & 1.6e-06 & 1.840(+110)(-90) & 1.030(+40)(-30) & CO & \citealt{Yang2025RAA} \\
J0348+0432 & 39.12 & 0.1024 & 2e-06 & 1.897(27) & 0.172(3) & He & \citealt{Antoniadis2013} \\
J1614$-$2230 & 3.15 & 8.687 & 1.333e-06 & 1.915(14) & 0.493(3) & boundary/CO & \citealt{Arzoumanian2018} \\
J0740+6620 & 2.89 & 4.767 & 5.07e-06 & 2.080(70) & 0.253(+6)(-5) & He & \citealt{Fonseca2021} \\
\hline
\multicolumn{8}{l}{\textit{Double neutron star systems}}\\
\hline
J1829+2456 & 41.01 & 1.176 & 0.1391 & 1.306(7) & 1.299(7) & \ldots & \citealt{Haniewicz2021} \\
B1534+12 & 37.90 & 0.4207 & 0.2737 & 1.333(2) & 1.345(2) & \ldots & \citealt{Fonseca2014} \\
J0737$-$3039A & 22.70 & 0.1022 & 0.08778 & 1.338(7) & 1.249(7) & \ldots & \citealt{Kramer2021} \\
J1756$-$2251 & 28.46 & 0.3196 & 0.1806 & 1.341(7) & 1.230(7) & \ldots & \citealt{Ferdman2014} \\
J1757$-$1854 & 21.50 & 0.1835 & 0.6058 & 1.341(3) & 1.392(3) & \ldots & \citealt{Cameron2023} \\
J0509+3801 & 76.54 & 0.3796 & 0.5864 & 1.399(6) & 1.412(6) & \ldots & \citealt{Martinez2024} \\
B1913+16 & 59.03 & 0.323 & 0.6171 & 1.438(1) & 1.390(1) & \ldots & \citealt{Weisberg2016} \\
J1518+4904 & 40.94 & 8.634 & 0.2495 & 1.470(32) & 1.248(+35)(-29) & \ldots & \citealt{Tang2024} \\
J0453+1559 & 45.78 & 4.072 & 0.1125 & 1.559(5) & 1.174(4) & \ldots & \citealt{Martinez2015} \\
J1913+1102 & 27.29 & 0.2062 & 0.08953 & 1.620(30) & 1.270(30) & \ldots & \citealt{Ferdman2020} \\
\hline
\multicolumn{8}{l}{\textit{Candidate double neutron star systems}}\\
\hline
J1906+0746 & 144.07 & 0.166 & 0.0853 & 1.291(11) & 1.322(11) & \ldots & \citealt{Desvignes2019} \\
B2127+11C & 30.53 & 0.3353 & 0.6814 & 1.358(10) & 1.354(10) & \ldots & \citealt{Jacoby2006} \\
J1807$-$2500B & 4.19 & 9.957 & 0.747 & 1.365(2) & 1.206(2) & \ldots & \citealt{Lynch2012} \\
\hline
\end{tabular}
\end{table*}



\bibliographystyle{mnras}
\bibliography{example} % if your bibtex file is called example.bib

% Don't change these lines
\bsp	% typesetting comment
\label{lastpage}
\end{document}